\UseRawInputEncoding
\documentclass[prd,twocolumn,showpacsl,amsmath,nofootinbib,floatfix,superscriptaddress]{revtex4}
\usepackage{graphicx,color,dcolumn,booktabs,bm,multirow}
\usepackage{longtable,lscape}
\usepackage{txfonts}
\usepackage{overpic}
\usepackage{amssymb}
\usepackage{indentfirst}
\usepackage{feynmf}   
\usepackage{slashed}  
\usepackage{cases}
\usepackage{color}
\usepackage{multirow}
\usepackage{epstopdf}
\usepackage{graphicx,color,dcolumn,booktabs,bm}
\usepackage{epstopdf}
\usepackage{ulem}

\usepackage[colorlinks,
            citecolor=green,
            anchorcolor=red,
            menucolor=red,
            linkcolor=red,
            filecolor=red,
            runcolor=red,
            urlcolor=blue,
            frenchlinks=red]{hyperref}

\newcommand{\sumint}{\kern 0.2 em {\textstyle\sum} \kern -1.1 em \int_X}

\begin{document}

\title{The electromagnetic form factors of $\Lambda_c$ hyperon in the vector meson dominance model}
\author{Junyao Wan}\affiliation{School of Physics, Southeast University, Nanjing
211189, China}
\author{Yongliang Yang}
\email{yangyl@qdu.edu.cn}
\affiliation{College of Physics, Qingdao University, Qingdao 266071, China}
\author{Zhun Lu}
\email{zhunlu@seu.edu.cn}
\affiliation{School of Physics, Southeast University, Nanjing 211189, China}

\begin{abstract}
We apply a modified vector meson dominance (VMD) model to analyze the electromagnetic form factors of the $\Lambda_c$ hyperon in the time-like reaction $e^+e^-\rightarrow \Lambda_c^+ \bar{\Lambda}_c^-$. In the model, we include the contributions from the vector charmed mesons and their excitations $\psi(1S)$, $\psi(2S)$, $\psi(3770)$, $\psi(4040)$, $\psi(4160)$ and $\psi(4415)$. We perform a combined fit to the available data on the Born cross section in reaction $e^+e^-\rightarrow \Lambda_c^+ \bar{\Lambda}_c^-$ and the ratio of the electromagnetic form factors $|G_E/G_M|$ to obtain the values of the model parameters. Our results show that the VMD model can simultaneously describe the data of electromagnetic form factors from the Belle and BESIII Collaborations, and the behavior of $|G_E/G_M|$ for the BESIII data can be qualitatively reproduced by the VMD model prediction at the threshold region. Moreover, we predict the single and double polarization observables in $e^+e^-\rightarrow \Lambda_c^+ \bar{\Lambda}_c^-$ reactions, which are experimentally accessible in the polarized process. We also obtain the form factors of the $\Lambda_c$ hyperon in the space-like region via analytic continuing the time-like form factors.
\end{abstract}

\maketitle
\section{introduction}

The electromagnetic form factors (EMFFs) $G_E$ and $G_M$ of hadrons are important physical quantities that encode the information of the perturbative and nonperturbative quantum chromodynamics (QCD) effects in hadrons~\cite{Pacetti:2015iqa,Cabibbo:1960zza,Cabibbo:1961sz}.
The time-like and space-like EMFFs of the proton and neutron have been extensively studied, e.g., in the $ep$ elastic scattering, $\bar{p}p$ annihilation and $e^+e^-$ annihilation processes~\cite{Pacetti:2015iqa,Ablikim:2015vga,Denig:2012by,Akhmetshin:2015ifg,Haidenbauer:2014kja,
Achasov:2014ncd,Lees:2013uta,Kuraev:2011vq,TomasiGustafsson:2005kc, Aubert:2005cb,Pedlar:2005sj,Brodsky:2003gs,Iachello:2004aq,Andreotti:2003bt,
Antonelli:1998fv,Bardin:1994am,Armstrong:1992wq}.
In the last two decades, the EMFFs of hyperons (e.g., $\Lambda$, $\Sigma$, $\Xi$) in the time-like region have also been investigated.
Particularly, the enhancement of the cross sections in reactions $e^+\, e^- \to Y \bar{Y}$ near threshold was measured and analyzed~\cite{Aubert:2007uf,Ablikim:2017tys,Ablikim:2017pyl,Ablikim:2019vaj,Haidenbauer:2020wyp,Ablikim:2020kqp,Li:2020lsb}.
It is found that the vector mesons and their excitations, as the intermediate states of reactions, play important roles in these processes.
That is, they could provide an explanation for the threshold enhancement effect of scattering cross section and large ratio between $G_E$ and $G_M$~\cite{Cao:2018kos,Yang:2019mzq,Lorenz:2015pba}.

In recent years, the EMFFs of charmed hyperon $\Lambda_c$ attracts a lot of interest both theoretically~\cite{Karliner:2006ny,Xie:2020dkm,Sharma:2013uka,Can:2013tna,Kim:2018nqf,Delpasand:2020vlb} and experimentally~\cite{Aubert:2005gt,Pakhlova:2008vn,Pal:2017ypp,Ablikim:2017lct,Kniehl:2020szu}, since the $\Lambda_c$ hyperon is the lightest baryon containing the charm quark.
Similar to the $\Lambda$ hyperon, the $\Lambda_c$ target is unfeasible and
the EMFFs of $\Lambda_c$ can not be accessed from exclusive experiments in the space-like region~\cite{Bisello:1990rf,Yang:2019mzq}.
On the other hand, the cross section of reaction $e^+ e^- \to \Lambda_ c^+ \bar{\Lambda}_c^-$ has been measured by the Belle and the BESIII Collaborations~\cite{Pakhlova:2008vn,Ablikim:2017lct}.
The ratio between the electric form factor $G_E$ and the magnetic form factor $G_M$ near threshold region is also available~\cite{Ablikim:2017lct}.
These measurements provide great opportunity to study the dynamics on the production of charmed baryon pairs and the time-like EMFFs of $\Lambda_c$.
Furthermore, by analytic continuation, the knowledge of EMFFs in the time-like region could be extended to study the EMFFs in the space-like region.

The VMD model has been recognized as a reliable theoretical approach in the study of the space-like electromagnetic form factors of hadrons.
It can describe the existing data of proton and neutron EMFFs in the space-like region quite well.
The approach was also extended to investigate the EMFFs of the $\Lambda$ hyperon~\cite{Yang:2019mzq} in the time-like region.
In the VMD model, the electromagnetic form factors receive contributions from two parts.
One is the intrinsic structure defined by the valence quarks, the other is the contribution form the meson clouds in terms of vector mesons.
Due to the isoscalar property of $\Lambda$, the contribution of $\rho$ meson and its excitations should be excluded.
In order to introduce a complex structure of EMFFs in time-like region, the decay widths of the vector mesons and their excitations are taken into account~\cite{OConnell:1995nse,Lorenz:2015pba,Yang:2019mzq}.
Particularly, the contributions from the excitations below the threshold of $\Lambda\bar\Lambda$ pair are involved.
The study shows that the inclusion of these excitations are essential to simultaneously describe the experiment data of the effective form factors, ratio $|G_E/G_M|$ and relative phase $\Delta\Phi$ for $\Lambda$ hyperon in a wide range of $\sqrt{s}$.

Encouraged by the success on the nucleon and $\Lambda$ hyperon, in this letter, we extend the VMD model to explore the EMFFs of the $\Lambda_c$ hyperon.
There is some difference between the VMD model for the $\Lambda$ hyperon and that for the $\Lambda_c$ hyperon.
Firstly, The production of the $\Lambda_ c^+ \bar{\Lambda}_c^-$ pair are related to $c\bar{c}$ pair, which has the quantum numbers $I^G(J^{PC})=0^-(1^{--})$.
Secondly, we neglect the contribution from $\omega$, $\phi$ and their excitations in this work, because the masses of $\omega$, $\phi$ and their excitations are far away from the threshold of $\Lambda_c\Bar\Lambda_c$.
Thus, we only take into account the contributions from the relevant charmonium excitations: $\psi(1S)$, $\psi(2S)$, $\psi(3770)$, $\psi(4040)$, $\psi(4160)$ and $\psi(4415)$.
Thirdly, the Coulomb final-state interactions should be considered, which is similar to the case of the proton~\cite{Lorenz:2015pba,Lichard:2018enc}.
Based on the above consideration, we can obtain the formula of the time-like form factors for $\Lambda_c$ by analytic continuation of the space-like form factors.

The remained content of the paper is organized as follows.
In Section \ref{Sec:2}, we present a detailed framework on the form factors of $\Lambda_c $ hyperon in the VMD model.
In Section \ref{Sec:3}, we fit the theoretical expressions for $G_E$ and $G_M$ to the experimental data of reaction $e^+ e^-\to \Lambda_c^+ \bar\Lambda_c^-$ from the Belle and the BESIII collaborations.
We also provide our predictions for the single and double polarization observables, the relative phase angle $\Delta\Phi$, as well as space-like form factors of $\Lambda_c$.
We summarize the paper in Section \ref{Sec:4}.

\section{Form factors of $\Lambda_c$ hyperon in the VMD model \label{Sec:2}}

\begin{figure}
  \centering
  \includegraphics[width=0.95\columnwidth]{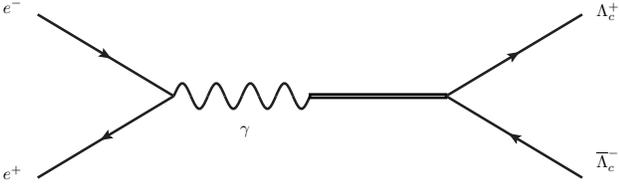}
  \caption{The reaction $e^+ e^-\to\Lambda_ c^+ \bar{\Lambda}_c^- $ depicted in the VMD model.}\label{FynVMD}
  \label{singlepolar}
\end{figure}

The process $e^+ e^- \to \Lambda_ c^+ \bar{\Lambda}_c^-$ which we study in the framework of the VMD model is shown in Fig.~\ref{FynVMD}.
That is, the photons formed in $e^+ e^-$ annihilation are first transformed into neutral vector mesons, the latter ones then decay into $\Lambda_ c^+ \bar{\Lambda}_c^-$ pairs through some specified couplings.
Since one-photon exchange dominates the production of spin $-1/2$ baryons $B$, the Born cross section of the process $e^+ e^- \to B \bar{B}$ can be parameterized~\cite{Cabibbo:1961sz} in terms of EMFFs.
Generally, the integrated cross section of the $\Lambda_c$ hyperon pairs production can be given in the following way:
\begin{align}
\sigma(s)={4\pi\alpha^2\beta\over 3s}C_{\Lambda_c}\bigg{[}|G_M(s)|^2+{1\over 2\tau}|G_E(s)|^2\bigg{]}\,
 \label{eq:crosssection}
\end{align}
Here, $G_E(s)$  and $G_M(s)$ are the electric form factor and the magnetic form factor in the time-like region, respectively, $\alpha$ is the fine-structure constant, $s$ is the square of the center of mass (c.m.) energy, $\tau={s/ 4M^2_{\Lambda_c}}$, and $\beta=\sqrt{1-1/\tau}$ is the velocity of the $\Lambda_c$ hyperon.
The Coulomb factor $C_{\Lambda_c}=\varepsilon~R$ parameterizes the electromagnetic interaction between the outgoing baryon and antibaryon, with $\varepsilon=\pi\alpha/\beta$ an enhancement factor resulting in a nonzero cross section at threshold and $R=1/(1-e^{-\pi\alpha/\beta})$ the Sommerfeld resummation factor~\cite{Sakharov:1948yq,Ablikim:2017lct}.

In the space-like region, the EMFFs $G_E (Q^2)$ and $G_M (Q^2)$ of $\Lambda_c$ hyperon can be expressed as
\begin{align}
G_M=F_1+F_2,~~~~~G_E=F_1-\tau F_2\,,
\label{eq:gegm}
\end{align}
where $\tau={Q^2/4M^2_{\Lambda_c}}$, and the $F_1(Q^2)$ and $F_2(Q^2)$ are the Dirac form factor and Pauli form factor respectively, which can be decomposed into
\begin{align}
F_i = F_i^S + F_i^V.
\end{align}
Here, $F_i^S$ and $F_i^V$ denote the isoscalar and isovector components of the form factors, respectively. Since the $\Lambda_c$ hyperon is an isospin singlet, the contribution from the isovector part $F_i^V$  should be excluded, which is similar to the case of the $\Lambda$ hyperon.
We note that the kinematic constraint $G_E(-4M_{\Lambda_c}^2)=G_M(-4M_{\Lambda_c}^2)$ is apparently satisfied in Eq.~(\ref{eq:gegm})~\cite{Iachello:2004ki}.

Previously, the VMD model have been widely used to study the EMFFs of the nucleon and the $\Lambda$ hyperon, showing that it has the advantage to well describe the experimental data in the space-like and the time-like regions~\cite{Bijker:2004yu,Bijker:2005cd,Bijker:2006kx,Iachello:2003ep,Iachello:2004aq,Yang:2019mzq}.
Very recently, it has also been applied to investigate the EMFFs for $\Sigma^+$ and $\Sigma^-$~\cite{Li:2020lsb}.
Encouraged by its success, we extend the model to study the EMFFs of the $\Lambda_c$ hyperon.
In the VMD model, two parts contributes to the Dirac form factor, one is the intrinsic structure, the other is the vector meson cloud;
while the Pauli form factor only receives the contribution from the meson cloud~\cite{Bijker:2006kx}.
As $\Lambda_c$ is charmed hyperon, we should consider the contribution of charmed mesons $J/\psi$ and their excitations.
Due to the isoscalar property of the $\Lambda_c$ hyperon, in principle the contributions of the vector mesons $\omega$, $\phi$ and their excitations should be included.
However, the masses of $\omega$, $\phi$ excitations are far low the threshold energy of $\Lambda_c\bar\Lambda_c$ compared to those of the charmed mesons, that is $|{m^2_{\omega(\phi)}\over m^2_{\omega(\phi)}-q^2}|\ll|{m^2_{J/\psi}\over m^2_{J/\psi}-q^2}|$ in VMD model.
Thus, in our modified model, we only investigate the charmed meson decay in the reaction $e^+e^-\rightarrow \Lambda_c\bar\Lambda_c$.
In this case, the contributing meson excitations are $\psi(1S)$, $\psi(2S)$, $\psi(3770)$, $\psi(4040)$, $\psi(4160)$, $\psi(4415)$.
We also assume that the expression of the form factors from the charmed excitations have the same form as those from the vector mesons $\omega$, $\phi$ and their excitations~\cite{Yang:2019mzq,Iachello:1972nu}, and we compare them to $\omega$ and $\phi$ in light of the magnitude of the mass of the charmed states, and put them into $\Omega$ and $\Phi$, namely $\omega(782)\rightarrow \psi(1S)$, $\phi(1020)\rightarrow \psi(2S)$ and so on.
The reason to take classification for $J/\psi$ and their excitations is to keep a fixed form for Dirac and Pauli form factors and simplify the structures in our modified model.
At $Q^2=0$, the EMFFs of $\Lambda_c$ hyperon can be normalized as follows,
\begin{align}
G_E(0)=1,~~~~~ G_M(0)=\mu_{\Lambda_c}\,,
\end{align}
where the magneton $\mu_{\Lambda_c}=0.48~\hat{\mu}_{N}$ is predicted in Ref.~\cite{Karliner:2006ny}.
One should note that result of the magnetic moment of the $\Lambda_c$ hyperon is given in unit of nucleon magneton.
Thus the magnetic moments with units of the $\Lambda_c$ hyperon natural magneton can be expressed as $\mu_{\Lambda_c}=1.039~\hat{\mu}_{\Lambda_c}$ using $\hat{\mu}_{\Lambda_c}={M_N\over M_c} \hat{\mu}_N$~\cite{Yang:2019mzq}.

Taking into account all the above constraints, we can write the parameterized forms of scalar parts of the Dirac and Pauli form factors in VMD model as follows:
\begin{eqnarray}
F_1^S(Q^2)&=&{g(Q^2)\over 3}\Sigma_{i=1}^N\bigg{[}1-\beta_{\Omega_i}-\beta_{\Phi_i}+\beta_{\Omega_i}
{m^2_{\Omega_i}\over m^2_{\Omega_i}+Q^2} \nonumber\\ &&+\beta_{\Phi_i}{m^2_{\Phi_i}\over m^2_{\Phi_i}+Q^2}\bigg{]}\,\label{F1s}\\
F_2^S(Q^2)&=&{g(Q^2)\over 3}\Sigma_{i=1}^N\bigg{[}(\mu_{\Lambda_c}-1-\alpha_{\Phi_i}){m^2_{\Omega_i}\over m^2_{\Omega_i}+Q^2}\nonumber\\ &&+\alpha_{\Phi_i}{m^2_{\Phi_i}\over m^2_{\Phi_i}+Q^2}\bigg{]}\,
\label{F2s}
\end{eqnarray}
where $N=3$. $\Omega_i \ (i=1,2,3)$ denotes the vector meson states $\psi(1S)$, $\psi(3770)$, and $\psi(4160)$, $\Phi_i\ (i=1,2,3)$ represents the vector meson states $\psi(2S)$, $\psi(4040)$, $\psi(4415)$.
Following Refs.~\cite{Iachello:1972nu,Bijker:2006kx,Yang:2019mzq}, the isoscalar part of $F_2$ appearing in Eq.~(\ref{F2s}) is constrained by the electric charges and magnetic moments of $\Lambda_c$, namely, $\alpha_{\Omega_i}=\mu_{\Lambda_c}-1-\alpha_{\Phi_i}$.
The intrinsic structure factor is a characteristic of valence quark structure and is chosen in a dipole form $g(Q^2)=(1+\gamma Q^2)^{-2}$, which is consistent with pQCD and fits the EMFFs of nucleon well~\cite{Bijker:2005cd,Bijker:2004yu,Iachello:1972nu}.
In the large $Q$ region, the forms also satisfy the constraints of the asymptotic behavior, $F_1 \sim 1/Q^4$ and $F_2 \sim 1/Q^6$. Furthermore, the coefficients $\beta_{\Omega_i}$, $\beta_{\Phi_i}$, $\alpha_{\Phi_i}$ can be naturally interpreted as the products of a $V\gamma$ coupling constant and a $VBB$ coupling constant~\cite{Iachello:1972nu}, respectively.
The parameter $\gamma$ in $g(Q^2)$ and the coefficients $\beta_{\Omega_i}$, $\beta_{\Phi_i}$, $\alpha_{\Phi_i}$ in Eqs.~(\ref{F1s})-(\ref{F2s}) are free parameters the values of which can be obtained by fitting the data of EMFFs.

By proper analytic continuation on the complex plane, we can obtain the form factors in the time-like region on the basis of the form factors in space-like region~\cite{TomasiGustafsson:2005kc,Iachello:2004aq}.
The analytic continuation in the time-like region is based on the following relation~\cite{TomasiGustafsson:2005kc}:
\begin{align}
Q^2= -q^2=q^2 e^{i\pi}\,.
\label{q2}
\end{align}
Therefore, in the time-like region, the intrinsic structure $g(q^2)$ has an analytical continuation form:
\begin{align}
 g(q^2)={1\over (1-\gamma q^2)^2}.
\end{align}
where $\gamma$ is a parameter larger than zero.
Thus, there is a pole in $g(q^2)$ in the position $q^2=1/\gamma$.
There are two methods to remove the pole, one is to change the relations in Eq.(\ref{q2}) with $Q^2\rightarrow q^2e^{i\theta} (\theta \neq \pi)$~\cite{Iachello:2004ki,Iachello:2004aq}, the
other is to impose the constraint $\gamma> 1/(4m^2_{\Lambda_c}) $ for the $\Lambda_c$ form factors~\cite{Yang:2019mzq}.
In this work, we will choose the latter one.
For the contribution of the meson cloud to the form factor, we take into account the widths of the vector charmed mesons $J/\psi$ and their excitations in order to introduce the complex structure of the EMFFs in the time-like region~\cite{Yang:2019mzq,Lorenz:2015pba}.
This lead to the following replacement in Eqs.~(\ref{F1s}) and (\ref{F2s})
\begin{eqnarray}
 {m^2_{\Omega_i}\over m^2_{\Omega_i}+Q^2} &\rightarrow& {m^2_{\Omega_i}\over m^2_{\Omega_i}-q^2-i m_{\Omega_i}\Gamma_{\Omega_i}}\,, \nonumber \\
 {m^2_{\Phi_i}\over m^2_{\Phi_i}+Q^2}&\rightarrow& {m^2_{\Phi_i}\over m^2_{\Phi_i}-q^2-i m_{\Phi_i}\Gamma_{\Phi_i}}\, .
 \label{Eq:Rep}
\end{eqnarray}
In this way we obtain the modified VMD model in the time-like region for $\Lambda_c$ hyperon. These terms are crucial for constructing the complex structure and reproducing the relative phase angle of the time-like EMFFs of $\Lambda_c$.
In principle, a momentum(energy)-dependent width~\cite{Kozyrev:2017agm} instead of a fixed width for each excitation of vector meson should be applied in Eq.~(\ref{Eq:Rep}).
However, as for the excited vector mesons included in the present work, there are more open channels, which makes their decay behaviors much complicated. Moreover, the decay widths and the branching ratios of the excitations of vector mesons are not well determined experimentally.
Thus, in the present work, we still use a fixed decay width as an approximation. We expect a fixed decay prescription will not change the result qualitatively.
Furthermore, the vector mesons can couple to such virtual mesons and these virtual mesons are supposed around $\Lambda_c$ hyperon. These contributions could also provide some momentum dependence to the form factors. In the present work, we do not include the virtual meson coupling in the VMD model.
As a matter of fact, it is very hard to consider such precise contribution in a relative rough mode, since there are so many undetermined coupling constants to describe the couplings of vector meson to the virtual mesons, such as the couplings related to excited vector meson with the pseudoscalar meson pair.
Such coupling indeed provides some effect of the meson cloud, thus it is possible to estimate the form factor of $\Lambda_c$ by considering the meson cloud coupling when the experimental data are abundant.

\section{Numerical results and discussions~\label{Sec:3}}

\subsection{Fit the time-like form factors}
\begin{figure}
  \centering
  \includegraphics[width=0.95\columnwidth]{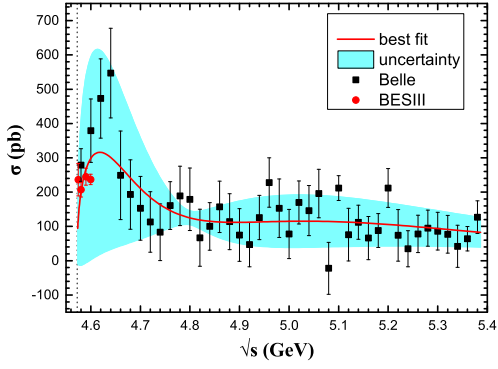}
  \caption{Our fit to the Born cross section $\sigma$ in reaction $e^+ e^-\to \Lambda_ c^+ \bar{\Lambda}_c^-$ . The rectangles and circles with error bars represent the data from the Belle~\cite{Pakhlova:2008vn} and BESIII~\cite{Ablikim:2017lct} Collaborations, respectively.}\label{BornCS}
\end{figure}

\begin{figure}
  \centering
  \includegraphics[width=0.95\columnwidth]{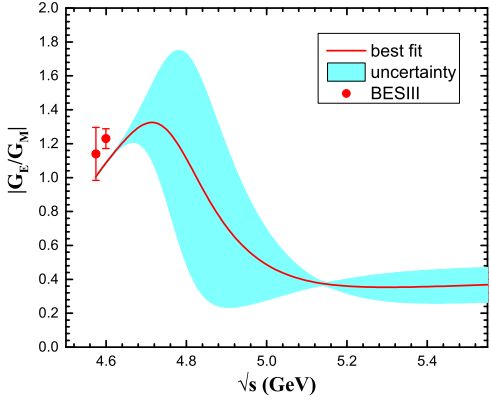}
  \caption{The same as Fig.~\ref{BornCS} but for the ratio $G_E/G_M$. The experimental data denote by the circles are from BESIII~\cite{Ablikim:2017lct} Collaboration. }\label{ratio}
\end{figure}

\begin{table}[htb]
\caption{The masses and widths of the involved charmed vector mesons in the model in unit of MeV \cite{Tanabashi:2018oca}. \label{Tab:Mass}}
\begin{tabular}{p{1.2cm}<{\centering}p{1.2cm}<{\centering}p{1.2cm}<{\centering}p{1.2cm}
<{\centering}p{1.2cm}<{\centering}p{1.2cm}<{\centering}}
\toprule[1pt]
State & Mass & Width &State &Mass & Width \\
\midrule[1pt]
$\psi(1S)$   &3097 & 0.093 & $\psi(2S)$  &3686 & 0.294 \\
$\psi(3770)$  &3773 &27.2 & $\psi(4040)$ &4039 & 80 \\
$\psi(4160)$  &4191 &70  &$\psi(4415)$  & 4421 & 62\\
\bottomrule[1pt]
\end{tabular}
\end{table}
We fit the expressions of the form factors in Eqs.~(\ref{F1s})-(\ref{F2s}) and the replacement in Eq.~(\ref{Eq:Rep}) to the experimental data of the Born scattering cross section and the EMFFs ratio  measured by the Belle~\cite{Pakhlova:2008vn} and BESIII~\cite{Ablikim:2017lct} Collaborations.
The data are in the range $4.59~\textrm{GeV} < \sqrt{s} <5.39~\textrm{GeV}$.
The masses and widths of the isoscalar vector mesons used in the fit are taken from Table~\ref{Tab:Mass}.
The values and the errors of the model parameters obtained from the fit are given in Table.~\ref{Tab:Parajpsi}, where the value of intrinsic parameter $\gamma=0.0899 \pm 0.0017~\textrm{GeV}^{-2}$, and the $\chi^2$ per degrees of freedom (d.o.f)  $\chi^2$ by $\Delta \chi^2 =1.436$.
The best fitted results are shown by the dashed lines in Figs.~\ref{BornCS} and \ref{ratio}.

It should be noted that $g(q^2)$ has a pole in the position $q^2=1/\gamma$, corresponding $q=3.335~\textrm{GeV}$ in our model.
Thus the poles of the intrinsic structure are restricted in the unphysical region and satisfied the constraint $\gamma>1/(4m_{\Lambda_c}^2)$ in this scenario.
Since the poles are below the threshold and we focus on the region above the $\Lambda_c^+ \bar{\Lambda}_c^-$ threshold, we can ignore the effect of the pole in the first place.

In Fig.~\ref{BornCS}, we show our fit to the Born cross sections in reaction $e^+ e^- \to \Lambda_c^+ \bar\Lambda_c^-$  measured by the Belle (filled square) and BESIII Collaboration (filled circle).
The vertical solid lines depict the threshold of $\Lambda_c^+ \bar{\Lambda}_c^-$.
We also provide the theoretical band corresponding to the uncertainty of parameters obtained from the errors of the data.
The Belle data cover the region $ 4.58~\textrm{GeV} < \sqrt{s}< 5.4 ~\textrm{GeV}$, while the BESIII data concentrate in the range $ 4.57~\textrm{GeV} < \sqrt{s}< 4.6 ~\textrm{GeV}$.
The comparison shows that the modified VMD model can describe the Belle and BESIII data after the theoretical error band is included.
It is found that the enhancement effect near the threshold of the $\Lambda_c^+ \bar\Lambda_c^-$ cross section can be well described in the VMD model.

In Fig.~\ref{ratio}, we present the model result of the ratio $|G_E/G_M|$ and compare it with the BESIII data (filled circle) which are near the threshold.
Again, the band corresponds to the uncertainty of parameters.It is shown that the prediction of VMD model do not contradict with the BESIII data of ratio. The model can qualitatively predict the behavior of ratio $|G_E/G_M|$.
It is worth noting that, according to the kinematic constraint, this ratio is equal to $1$ at the threshold of $\Lambda_c^+ \bar\Lambda_c^-$, which is an important constraint for the form factors in the time-like region.
Furthermore, the ratio increases with increasing $\sqrt{s}$ in the near threshold region and reaches the maximum value 1.3 at around $\sqrt{s}=4.7$ GeV.
This trend is also consistent with the BESIII data.
In the larger $\sqrt{s}$ region, the ratio decreases with increasing $\sqrt{s}$.
These features are similar to those of the $\Lambda$ hyperon~\cite{Yang:2019mzq,Haidenbauer:2016won}.
Due the VMD model, the asymptotic behaviors of form factors, the ratio of EMFFs satisfy a fact that the result tends to be constant in the limit of $q^2\to\infty$~\cite{Yang:2019mzq}.

\begin{table*}[htb]
\caption{The values and the errors of the parameters obtained from the fit. \label{Tab:Parajpsi}}
\begin{tabular}{p{1.5cm}<{\centering} p{2.2cm}<{\centering} p{1.5cm}<{\centering} p{2.2cm}<{\centering} p{1.5cm}<{\centering} p{2.2cm}<{\centering}}
\toprule[1pt]
Parameter & Value & Parameter &Value & Parameter &Value \\
\midrule[1pt]
$\beta_{\psi(1S)}$ & $7.9636\pm 0.3507$  &
$\beta_{\psi(3770)}$ & $-2.1338\pm 0.1746$  &
$\beta_{\psi(4160)}$ & $-1.2008\pm 0.1092$ \\
$\beta_{\psi(2S)}$ &  $-2.1857\pm 0.1839$ &
$\beta_{\psi(4040)}$ & $-1.6917\pm0.1529$  &
$\beta_{\psi(4415)}$ & $-1.0592\pm0.0597$\\
$\alpha_{\psi(2S)}$ & $-4.5310 \pm 0.1107$  &
$\alpha_{\psi(4040)}$ & $23.2260\pm0.7772$  &
$\alpha_{\psi(4415)}$ & $0.6651\pm0.0934$ \\
\bottomrule[1pt]
\end{tabular}
\end{table*}

\begin{figure}
\centering
  \includegraphics[width=0.95\columnwidth]{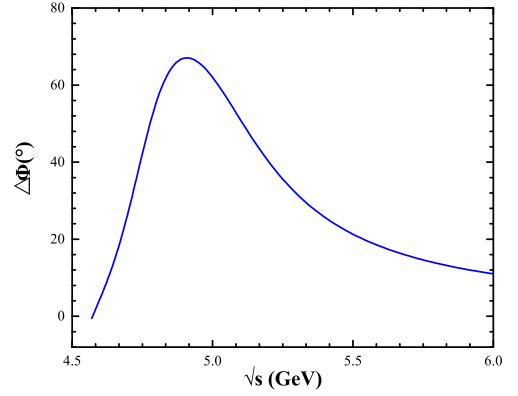}
  \caption{The prediction for the relative phase $\Delta \Phi$ vs $\sqrt{s}$ in $e^+ e^- \rightarrow \Lambda_c^+ \bar{\Lambda}_c^-$ with parameters in Table ~\ref{Tab:Parajpsi}. }\label{arg}
\end{figure}

As $G_E$ and $G_M$ in the time-like region are complex, there is a relative phase $\Delta \Phi$ between them.
The measurement of this phase at different $\sqrt{s}$ could provide additional information of EMFFs which can not been revealed by $|G_E|$ and $|G_M|$.
Using the values of the parameter extracted from the Belle and BESIII data, we predict the relative phase of the EMFFs of the $\Lambda_c$ as function of $\sqrt{s}$, as shown in Fig.~\ref{arg}.
One should note that $\Delta \Phi=0$ at the $\Lambda_ c^+ \bar{\Lambda}_c^-$ threshold due to $G_E=G_M$ at $s=4M_{\Lambda_c}^2$.

\subsection{Prediction of polarization observables in time-like region}

\begin{figure}[htb]
  \centering
  \includegraphics[width=0.95\columnwidth]{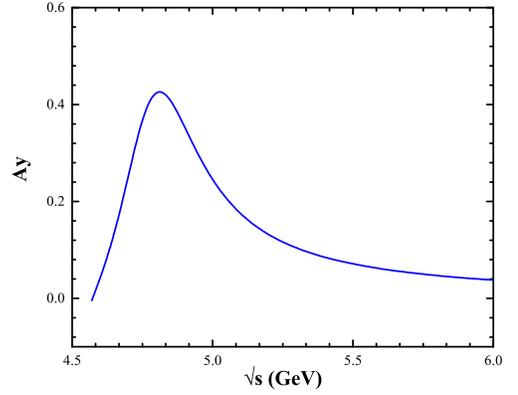}
  \caption{Similar to Fig.~\ref{arg}, but for the single polarization observable $A_y$ vs $\sqrt{s}$ in $e^+ e^- \rightarrow \Lambda_c^+ \bar{\Lambda}_c^-$ at the fixed angle $\theta=45^\circ$.}
  \label{Ay}
\end{figure}
\begin{figure*}[htb]
  \centering
  \includegraphics[width=0.99\columnwidth]{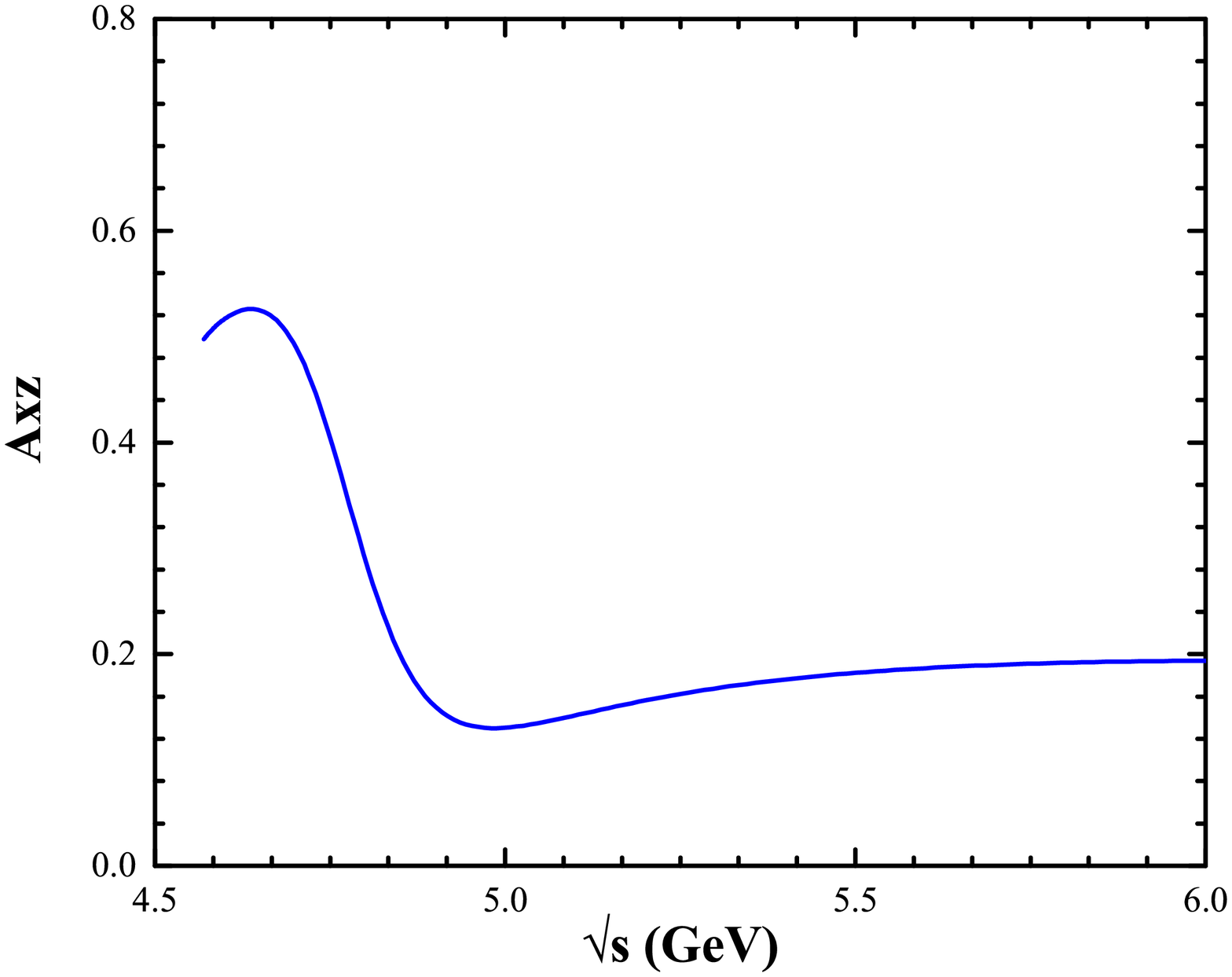}~~~~~~
  \includegraphics[width=0.99\columnwidth]{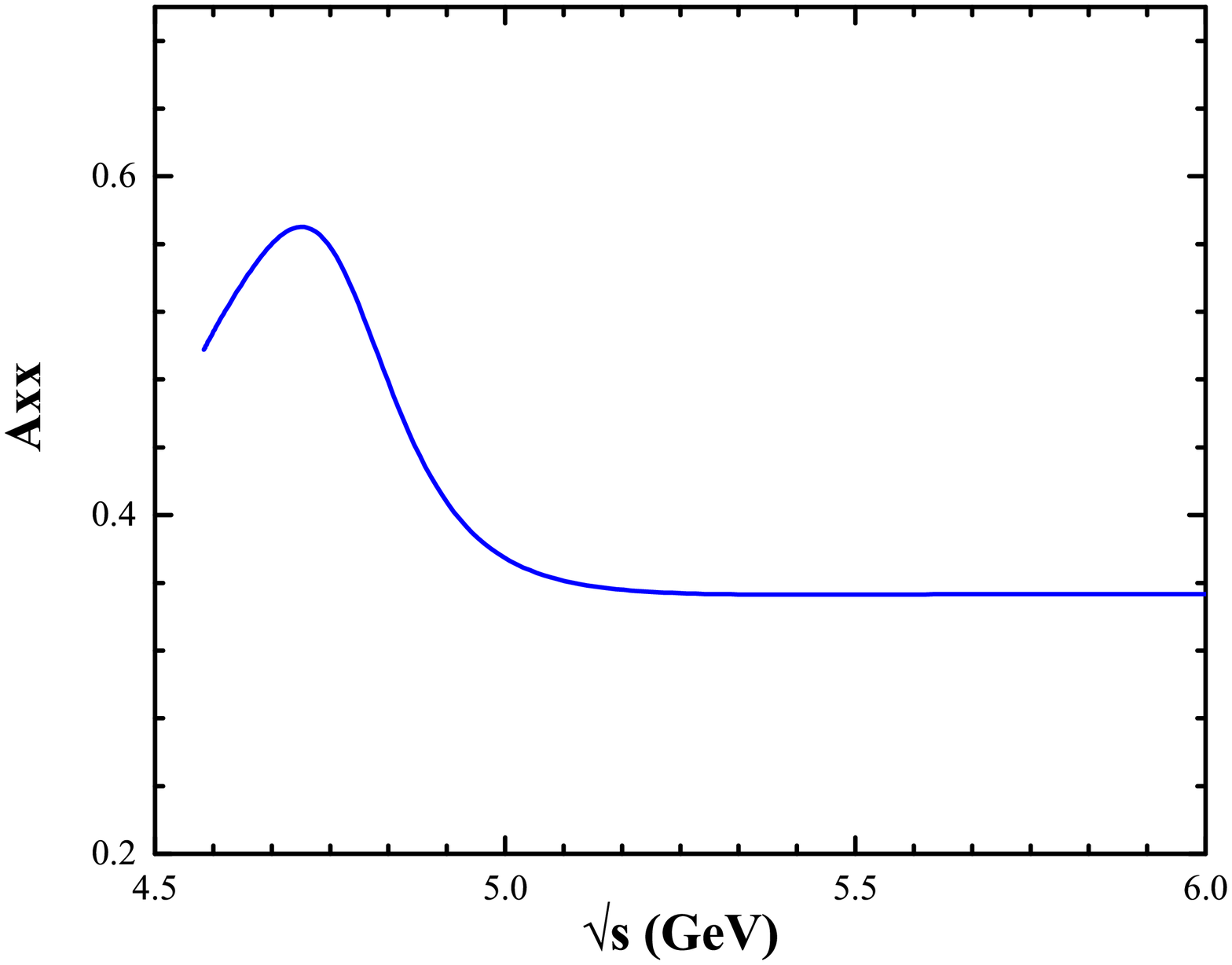}\\
  \includegraphics[width=0.99\columnwidth]{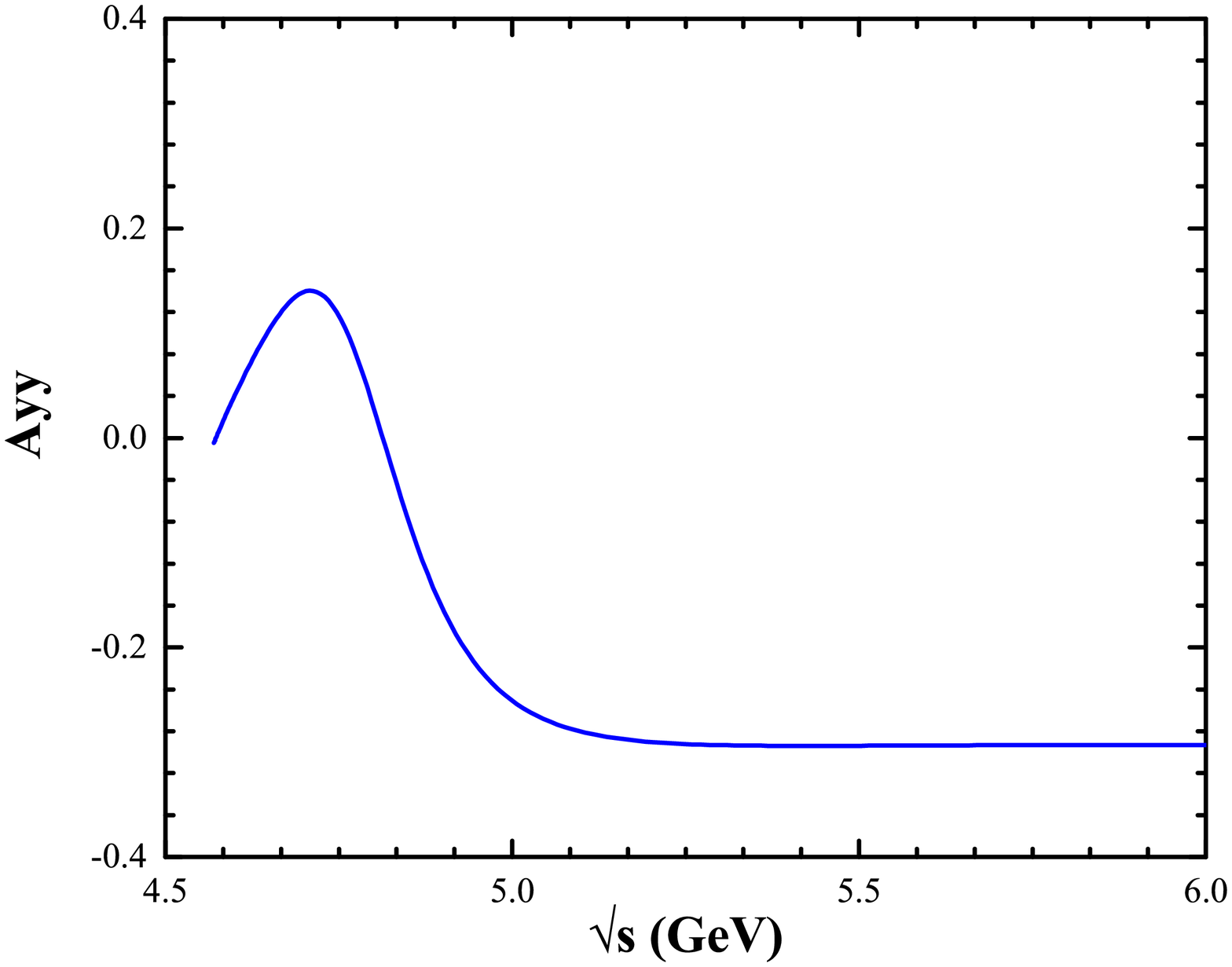}~~~~~~
  \includegraphics[width=0.99\columnwidth]{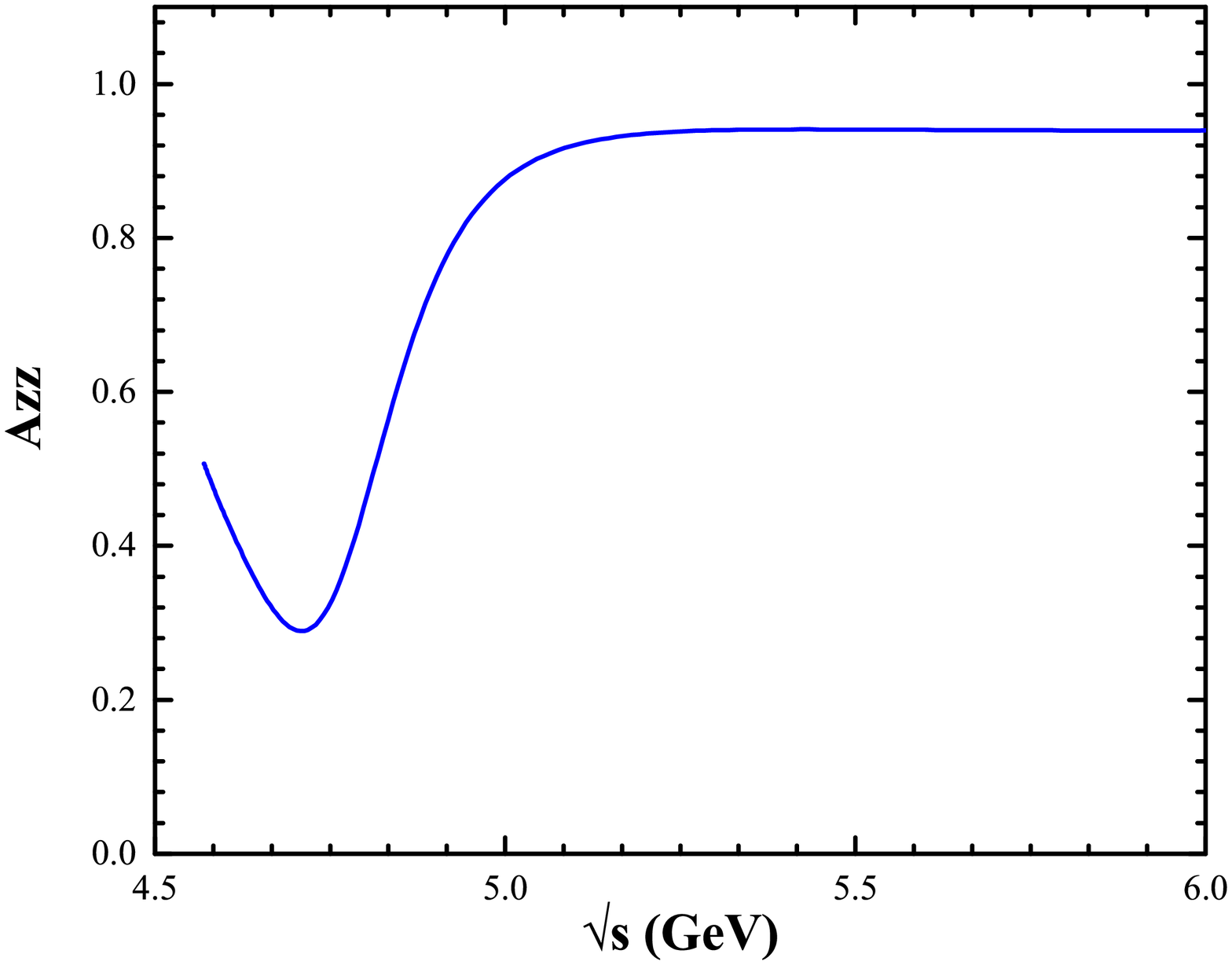}
  \caption{Similar to Fig.~\ref{arg} but for the double polarization observables $A_{xz}$, $A_{xx}$, $A_{yy}$ and $A_{zz}$ vs $\sqrt{s}$ in $e^+ e^- \rightarrow \Lambda_c^+ \bar{\Lambda}_c^-$ at the fixed angle $\theta=45^\circ$.}
  \label{Axz}
\end{figure*}

In the case of final $\Lambda_c\bar\Lambda_c$ with polarizations, the corresponding expressions for the cross section have been given in Refs.~\cite{TomasiGustafsson:2005kc,Faldt:2016qee,Faldt:2017kgy}.
The single and double polarization observables are defined by
$A_{i}=\bigl({d\sigma\over d\Omega_{\Lambda_c}}\bigr)_{i}/\bigl({d\sigma\over d\Omega_{\Lambda_c}}\bigr)_0$ and $A_{ij}=\bigl({d\sigma\over d\Omega_{\Lambda_c}}\bigr)_{ij}/\bigl({d\sigma\over d\Omega_{\Lambda_c}}\bigr)_0$, where $\bigl({d\sigma\over d\Omega_{\Lambda_c}}\bigr)_0$ is unpolarized cross section, and $\bigl({d\sigma\over d\Omega_{\Lambda_c}}\bigr)_{i(j)} (i, j=x, y, z)$ correspond to the polarized cross section.
Using the model results of $G_E$ and $G_M$, we present the prediction for the spin-dependent observables in reaction $e^+ e^- \rightarrow \Lambda_c^+ \bar{\Lambda}_c^-$.
In the single-photon exchange approximation, the single and double spin polarization observables can be expressed in terms of $G_E$ and $G_M$~\cite{Faldt:2016qee,Faldt:2017kgy}:
\begin{align}\label{eq:doublespin3}
A_{y}=&{-2M_\Lambda\sqrt{s}\sin(2\theta)~\rm{Im}(G_MG_E^*)\over{D_c-D_s\sin^2(\theta)}}
\,,\nonumber\displaybreak[0] \\
A_{xz}=&{2M_\Lambda\sqrt{s}\sin(2\theta)\rm{Re}(G_MG_E^*)\over{D_c-D_s\sin^2(\theta)}}
\,, \nonumber \displaybreak[0]\\
A_{xx}=&{[D_c-D_s]\sin^2(\theta)\over{D_c-D_s\sin^2(\theta)}}\,, \nonumber \displaybreak[0]\\
A_{yy}=&{-D_s\sin^2(\theta)\over{D_c-D_s\sin^2(\theta)}}\,, \nonumber \displaybreak[0]\\
A_{zz}=&{[D_s\sin^2(\theta)+D_c\cos^2(\theta)]\over{D_c-D_s\sin^2(\theta)}}\,,
\end{align}
where $\theta$ is the scattering angle defined in the c.m.frame,
and $D_c=2s|G_M|^2$, $D_s=s|G_M|^2-4M^2|G_E|^2$.
In Fig.~\ref{Ay}, we present our numerical result of the single polarization observable $A_y$ vs $\sqrt{s}$, which depends on the imaginary part of the product of $G_MG_E^*$.
As a demonstration, in the calculation we fixed the scattering angle $\theta=45^\circ$.
The prediction shows that the shape of $A_y$ is similar to that of the relative phase $\Delta \Phi$ in Fig.~\ref{arg}, since $\rm{Im}(G_M G_E^*)$ is proportional to $\sin(\Delta\Phi)$.
This indicates that exact information of $\Delta \Phi$ could be obtained from the precise measurement of the single spin polarization $A_y$.
In addition, we plot the double polarization observables $A_{xz}$, $A_{xx}$, $A_{yy}$ and $A_{zz}$ vs $\sqrt{s}$ in Fig.~\ref{Axz}.
It is found that in the near threshold region, the polarization observables changes drastically with $\sqrt{s}$, while in the large $\sqrt{s}$ region, the double polarization observables almost remain unchanged.

\subsection{Form factors in space-like region}

\begin{figure*}[htb]
  \centering
  \includegraphics[width=0.95\columnwidth]{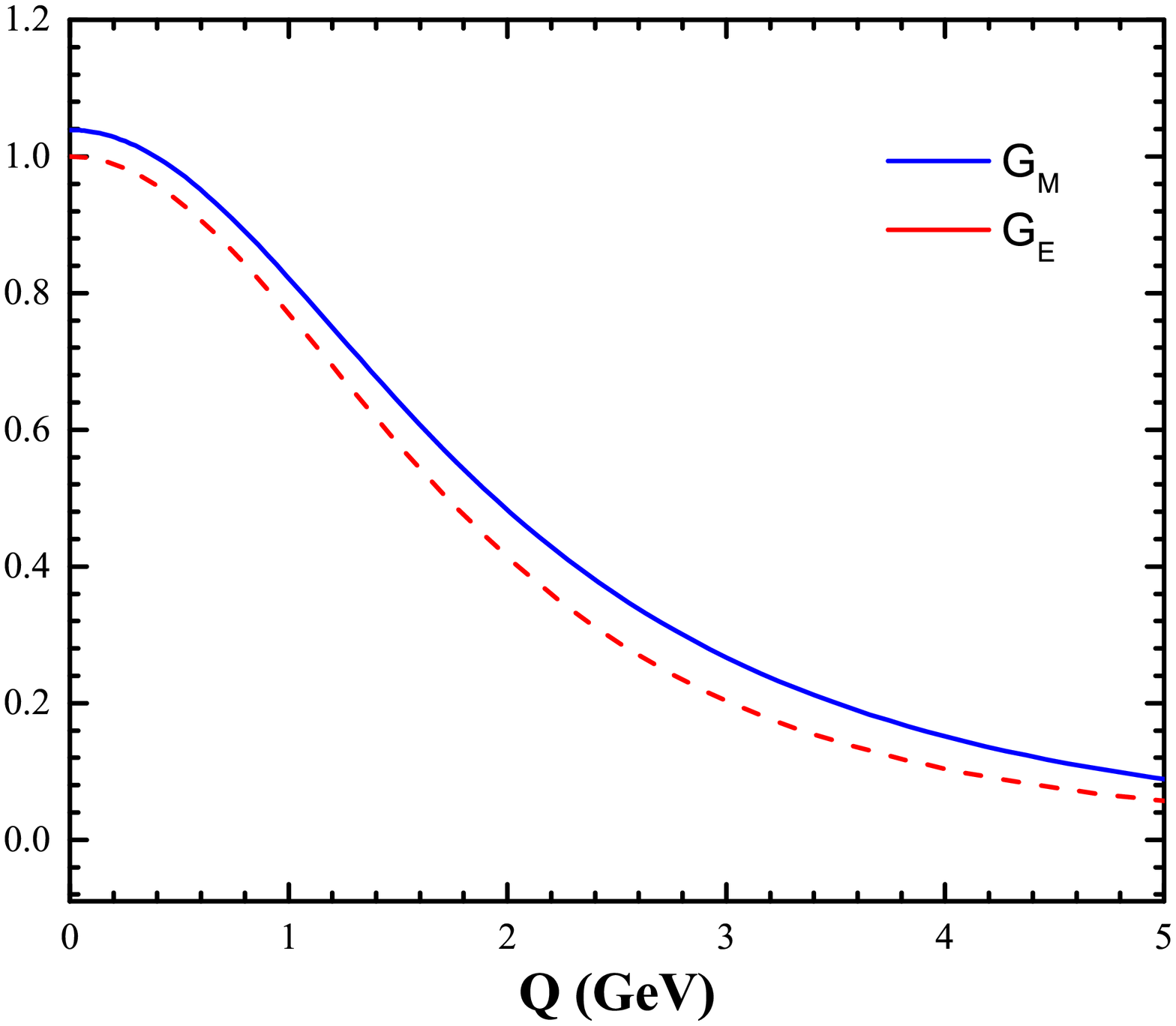}
  \includegraphics[width=0.95\columnwidth]{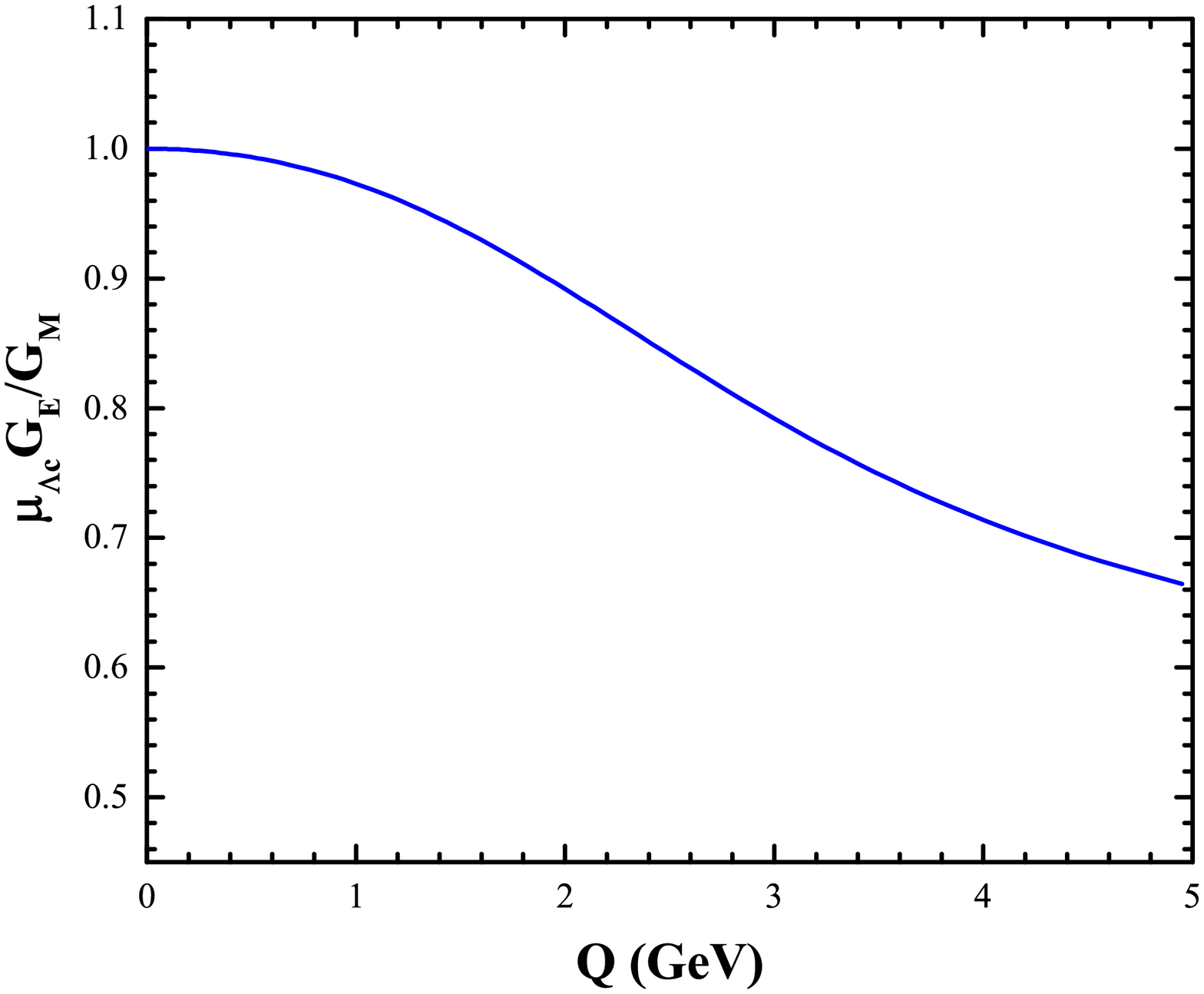}
  \caption{The EMFFs of the $\Lambda_c$ hyperon from the present estimation in the space-like region. The solid curve and the dotted curve in the left panel depict $G_E$ and $G_M$, respectively.
  The solid curve in the right panel is the ratio $\mu_{\Lambda_c}G_E/G_M$.}
  \label{gmge}
\end{figure*}

The EMFFs of $\Lambda_c$ in the space-like region can be directly calculated using Eqs.~(\ref{F1s}), (\ref{F2s}) and the model parameters in Table~\ref{Tab:Parajpsi}.
We perform the numerical calculation on the space-like $G_M$ and $G_E$ of $\Lambda_c$ vs $Q^2$ and present the results in the left panel of Fig.~\ref{gmge}, which shows that the magnitude and the shape of $G_E$ are similar to that of $G_M$.
A more clear picture about the relative size of $G_E$ and $G_M$ can be revealed by the ratio $\mu_{\Lambda_c}G_E/G_M$, as depicted in the right panel of Fig.~\ref{gmge}.
It is found that the ratio is smaller than 1 and decrease with increasing $Q^2$ in space-like region, which is similar to the case of the proton~\cite{TomasiGustafsson:2005kc,Punjabi:2015bba}.
However, it is larger than the ratio of the proton EMFFs.

\section{Summary \label{Sec:4}}

In this letter, we have investigated both the time-like and space-like EMFFs of $\Lambda_c$ using a modified VMD model.
In this model, the EMFFs are contributed by two parts. One is the intrinsic structure part, the other is the meson clouds part.
Similar to the $\Lambda$ hyperon, the contributions from the isovector components to the Dirac and Pauli form factors vanish due to the isoscalar property of the $\Lambda_c$ hyperon.
We have taken into account the contributions from the isospin-singlet charmed vector mesons $\psi(1S)$, $\psi(2S)$, $\psi(3770)$, $\psi(4040)$, $\psi(4160)$, $\psi(4415)$.
The inclusion of these mesons and their widths can naturally produce the complex structure of the time-like EMFFs.
Using the VMD model expressions for the EMFFs of the $\Lambda_c$, we have fit the Born cross section in reaction $e^+ e^- \to \Lambda_c^+\bar\Lambda_c^-$ to the data from the Belle and BESIII experiments.
We have also included the ratio $|G_E/G_M|$ measured by BESIII.
We find that the modified VMD model can describe the Born cross section of $e^+ e^- \to \Lambda_c^+\bar{\Lambda}_c^-$  at Belle and BESIII simultaneously.
Particularly, the enhancement effect near the threshold of the $\Lambda_c^+ \bar\Lambda_c^-$ pair can be described in the VMD model.
It is also shown that the VMD model can qualitatively describe the ratio $|G_E/G_M|$.
We have presented the numerical results of the relative phase $\Delta\Phi$, and predicted the single and double polarization observables in the process $e^+e^-\rightarrow \Lambda_c^+ \bar{\Lambda}_c^-$.
The measurement of these quantities could be used to verify the validity of the model.
Finally, we have extended the time-like EMFFs to space-like region using the parameter values obtained from the fit.
The numerical results show that the magnitudes and shapes of $G_E$ and $G_M$ is rather similar.

\section{Acknowledgements}
We would like to acknowledge useful discussions with Xiao-Rui Lyu. This work is partially supported by the National Natural Science Foundation of China (No.~11575043) and Shandong Provincial Natural Science Foundation, China (Grants No. ZR2020QA081).

\end{document}